\documentstyle[12pt]{article}
\def\al{\alpha}
\def\ga{\gamma}
\def\de{\delta}
\def\th{\theta}
\def\ka{\kappa}

\def\si{\sigma}
\def\ps{\psi}
\def\De{\Delta}

\def\bea{\begin{eqnarray}}
\def\eea{\end{eqnarray}}
\def\etal{{\it et al.}}

\def\fr#1#2{{{#1} \over {#2}}}
\def\ket#1{|{#1}\rangle}
\def\half{{\textstyle{1\over 2}}}
\def\lsim{\mathrel{\rlap{\lower4pt\hbox{\hskip1pt$\sim$}}
    \raise1pt\hbox{$<$}}}

\def\hydrogen{H}
\def\antihydrogen{$\overline{\rm{H}}$}

\newcommand{\beq}{\begin{equation}}
\newcommand{\eeq}{\end{equation}}
\newcommand{\rf}[1]{(\ref{#1})}

\renewenvironment{thebibliography}[1]
 { \rm
   \begin{list}{\arabic{enumi}.}
    {\usecounter{enumi} \setlength{\parsep}{0pt}
     \setlength{\itemsep}{3pt} \settowidth{\labelwidth}{#1.}
     \sloppy
    }}{\end{list}}

\begin{document}
\begin{center}
{{\large\bf Investigating Lorentz and CPT Symmetry with
Antihydrogen\footnote{Invited talk at the XIVth Rencontres de Blois on
Matter-Antimatter Asymmetry,
Chateau de Blois, France, 16-22 June 2002}\\} \vspace{0.2cm}
Neil Russell\\
{\small\it Physics Department, Northern Michigan University,
Marquette, MI 49855, U.S.A.\\
email: nrussell@nmu.edu\\}}
\vspace{0.3cm}
\parbox{6in}{\small
This talk discusses theoretical aspects of tests of
CPT and Lorentz Symmetry that will in principle
be possible with trapped antihydrogen.
The framework is the standard-model extension,
which admits minuscule violations of CPT and Lorentz symmetry
in a general manner
without giving up other features of the standard model
of particle physics.
Spectroscopic transitions in hydrogen and antihydrogen
that exhibit leading-order effects are identified.
Such comparisons of spectral frequencies in antimatter
with the corresponding frequencies in regular matter
will bound parameter combinations
that are not accessible with regular-matter atoms alone.}
\end{center}

\vspace{.4cm}\noindent
{\bf 1. The Standard-Model Extension}\\[.2cm]
The laws of physics appear to be unchanged
under the Lorentz and CPT transformations.
These symmetries are built into the standard model
of particle physics,
and so calculations in the usual standard-model context
cannot predict how
violations of these symmetries might occur.
However, the standard-model extension \cite{ck}
allows symmetry-violating effects
to be investigated in an explicit microscopic framework.
In this proceedings,
I give an overview of results in this context,
mention the experimental findings,
and then focus on hydrogen-antihydrogen symmetry tests
investigated in collaboration with
R.\ Bluhm and V.A.\ Kosteleck\'y \cite{bkrHbar}.

It is possible that
a Lorentz- and CPT-symmetric fundamental theory
underlying the standard model
could violate these symmetries \cite{qqcpt01},
perhaps through spontaneous symmetry breaking
at lower energies \cite{qqkps}.
No satisfactory fundamental theory is known,
but a candidate is string theory.
A practical way to proceed is to perform
calculations in the context of the
standard-model extension \cite{ck}.
The standard-model extension adds to the conventional
standard model all possible terms built from conventional fields
that violate Lorentz symmetry while
preserving observer coordinate independence.
This involves a large set of parameters,
which can be reduced by making simplifying assumptions.
For example, it is often useful to require that the terms
be gauge invariant, position independent, and renormalizable.
Since the standard-model extension is expected to be
the low-energy limit of an underlying theory,
the symmetry violations would most likely be suppressed
by ratios involving the low-energy mass and
the Planck mass.
Other theoretical issues about the standard-model extension
include supersymmetry \cite{qqbek},
noncommutative field theory \cite{qqchklo},
and anomaly cancellations \cite{qqjk}.

The standard-model extension is applicable
to all areas of physics.
Applications of this
framework include a possible mechanism for generating
the baryon asymmetry in the universe \cite{qqbckp}.
For the neutral mesons,
the prediction of sidereal dependencies \cite{qqak}
in transition probabilities
have been tested in the neutral kaons \cite{qqkexpt}
and in the neutral-B systems \cite{qqbexpt}.
Other relevant aspects
of neutral-meson physics
include efforts to bound symmetry parameters
in the neutral-D system \cite{qqckpvi}
and studies of analogue models \cite{kr}.
In the photon sector,
excellent bounds on Lorentz symmetry
have been placed through analysis of
light from distant cosmological
sources \cite{ck,qqkm}.
In the lepton sector,
recent results on muonium
and on anomaly-frequency comparisons
of oppositely-charged muons
at CERN and BNL have been obtained \cite{qqvh}.
Electron-positron comparisons using Penning traps
have placed several bounds on standard-model extension
parameters \cite{qqeexpt}.
An experiment with a spin-polarized torsion pendulum
has obtained impressive results \cite{qqeexpt2}.

Clock-comparison experiments \cite{qqccexpt},
comparing spectral lines of atomic transitions,
are capable of Planck-scale tests
of Lorentz symmetry \cite{qqkla}.
In the context of the standard-model extension,
tests have been performed with
Hg and Cs magnetometers \cite{qqlh},
a noble-gas maser \cite{qqdb},
and with an H maser \cite{qqdp}.
In such experiments,
rotational symmetry is tested
by monitoring frequency variations of
a Zeeman hyperfine transition as the quantization axis
of the clock changes direction
with the Earth's rotation.
A practical way to do this is to operate two atomic clocks
of different atomic species in the same laboratory,
so as to avoid issues with signal propagation.
To increase the sensitivity of clock-comparison
experiments to other components of the
standard-model extension couplings,
and to address various other limitations,
there is considerable interest in mounting
experiments of this type on the International Space Station
or other spacecraft \cite{qqspaceexpt}.

Calculations in the standard-model extension context
indicate that comparisons of
matter and antimatter counterparts
provide clean tests of CPT symmetry.
It would be ideal to compare the narrow spectral lines of an atomic clock
with the spectral lines of the corresponding antimatter clock.
However, high-Z atoms like Rb or Cs used in current clocks
are unlikely to be produced in antimatter form.
For atoms of low atomic number, however,
developments indicate that it may soon be possible
to investigate the spectral lines of cold antimatter.

\vspace{.4cm}\noindent
{\bf 2. Antihydrogen}\\[.2cm]
After experiments at CERN and Fermilab produced
hot free antihydrogen (\antihydrogen) \cite{hotHbar},
the ATHENA and ATRAP
collaborations \cite{collabs} at CERN
have worked towards the goals of obtaining
cold and trapped antihydrogen \cite{hbargenl}.
Numerous technological advances have
been made along the way \cite{hbarstatus}.
The reported production of cold \antihydrogen\
atoms \cite{coldHbar}
by the ATHENA collaboration
marks a milestone on this path.
When spectroscopy of \antihydrogen\ atoms is performed,
it will mark the first direct comparisons of neutral atoms
with the corresponding antimatter atoms,
and clean tests of CPT symmetry are likely to result.

From the experimental perspective,
a comparison of the 1S to 2S transition in
the two systems would be natural
because of the
parts in 10$^{14}$ and 10$^{12}$
precisions possible
with hydrogen (H) beams \cite{hansch}
and with trapped \hydrogen\ \cite{cesar}.
In principle,
the maximum possible precision
is about a part in 10$^{18}$,
or about a 1-mHz resolution \cite{hanschICAP}.
With the same precision in \antihydrogen,
an excellent test of CPT would be achieved.

From the theoretical perspective,
there are many possible transitions in
\hydrogen\ and \antihydrogen\ atoms,
and the challenge is to identify those
that are most likely to show CPT-violating
differences.
The point is that some will be suppressed by
powers of the fine structure constant,
or perhaps by the speed of the atoms in
the reference frame.
The standard-model extension provides
a calculational framework of powerful generality
for analyzing the various transitions in \hydrogen\ and \antihydrogen.
In the following sections,
I consider the spectroscopy of free atoms,
and then atoms confined in magnetic traps.
The 1S to 2S transitions and the hyperfine transitions are considered.

\vspace{.4cm}\noindent
{\bf 3. Free Hydrogen and Antihydrogen}\\[.2cm]
The effect of the standard-model extension
on free atoms of \hydrogen\ or \antihydrogen\
can be studied by calculating the
perturbative shifts in
energy levels of the quantum stationary states,
starting from the modified Dirac equation,
\beq
\left( i \ga^\mu D_\mu - m_e - a_\mu^e \ga^\mu
- b_\mu^e \ga_5 \ga^\mu
- \half H_{\mu \nu}^e \si^{\mu \nu}
+ i c_{\mu \nu}^e \ga^\mu D^\nu
+ i d_{\mu \nu}^e \ga_5 \ga^\mu D^\nu \right) \ps = 0
\quad .
\label{dirac}
\eeq
In this expression,
$\ps$ is a four-component electron field describing
an electron of mass $m_e$ and
charge $q = -|e|$.
We define $i D_\mu \equiv i \partial_\mu - q A_\mu$,
and use units with $\hbar = c = 1$.
The field interacts with the proton Coulomb potential
$A^\mu = (|e|/4 \pi r, 0)$.
The coefficients
$a_\mu^e$, $b_\mu^e$, $H_{\mu \nu}^e$, $c_{\mu \nu}^e$
and $d_{\mu \nu}^e$
parameterize Lorentz violation,
and the coefficients
$a_\mu^e$ and $b_\mu^e$
also parameterize CPT violation.
All are expected to be minuscule \cite{ck}.
Energy shifts in the \hydrogen\ and \antihydrogen\ systems
could also be due to perturbative couplings
$a_\mu^p$, $b_\mu^p$, $H_{\mu \nu}^p$, $c_{\mu \nu}^p$,
and $d_{\mu \nu}^p$
in a modified Dirac equation for free protons.

In conventional
relativistic quantum mechanics,
the \hydrogen\ and \antihydrogen\ hamiltonians
are identical:
reversing the charges of the particles
leaves the interaction potential unchanged,
and the reduced mass is also unaffected.
Thus, the eigenfunctions and associated energies
are the same for \hydrogen\ and \antihydrogen,
and within conventional physics
there should be no
observable differences between the spectra of
\hydrogen\ and \antihydrogen.
This argument carries over to
all perturbative effects
from conventional quantum electrodynamics.

If the Lorentz- and CPT-violating couplings
are included in a perturbative calculation,
the energy shifts in the spectrum of
the \hydrogen\ electron can differ from
those for the \antihydrogen\ positron.
To make the comparison,
the perturbative effect in \antihydrogen\
is found from Eq.\ \rf{dirac}
by a method involving charge conjugation and
appropriate field redefinitions \cite{bkr}.
The energy perturbations arising from the
proton and antiproton may also be included
by using relativistic two-fermion
techniques \cite{D2}.
In the uncoupled basis with electronic angular momentum $J$
and nuclear angular momentum $I$,
and corresponding third components $m_J$, $m_I$,
the perturbative Lorentz-violating energy corrections
for the basis states $\ket{m_J,m_I}$
may be found.
With this approach,
the energy corrections for protons or antiprotons
have the same mathematical form as those
for electrons or positrons,
except for the replacement of superscripts $e$ with $p$.
The shifts in the 1S and 2S energy levels
for free \hydrogen\
are found to be matched at leading order:
\bea
\De E^{H} (m_J, m_I)
& \approx &
(a_0^e + a_0^p - c_{00}^e m_e - c_{00}^p m_p)
\cr
&&
+ (-b_3^e + d_{30}^e m_e + H_{12}^e) {m_J}/{|m_J|}
\cr
&&
+ (-b_3^p + d_{30}^p m_p + H_{12}^p) {m_I}/{|m_I|} ~ ,
\label{EHJI}
\eea
where $m_p$ is the proton mass.
The same is true for \antihydrogen;
the leading-order energy shifts
$\De E^{ \overline{H}}$
in the 1S and 2S levels
are identical.
They are obtained from the expression \rf{EHJI}
with the substitutions
$a_\mu^e \rightarrow - a_\mu^e$,
$d_{\mu \nu}^e \rightarrow - d_{\mu \nu}^e$,
$H_{\mu \nu}^e \rightarrow - H_{\mu \nu}^e$;
$a_\mu^p \rightarrow - a_\mu^p$,
$d_{\mu \nu}^p \rightarrow - d_{\mu \nu}^p$,
$H_{\mu \nu}^p \rightarrow - H_{\mu \nu}^p$.

To find the effect of the Lorentz-violation on the
frequency spectrum of \hydrogen\ and \antihydrogen,
we must calculate appropriate energy differences.
Since the electron and proton spins in \hydrogen\
interact through the hyperfine interaction,
the appropriate basis is the coupled basis
$\ket{F,m_F}$,
where $F$ is the total angular momentum,
and $m_F$ is the third component.
Similarly, this is true for the
positron and antiproton spins
in \antihydrogen.
There are four allowed 1S--2S two-photon transitions
in \hydrogen\ and \antihydrogen,
determined by the selection rules
$\De F = 0$ and $\De m_F = 0$ \cite{cagnac}.
In each, the spins remain unchanged.
It follows from \rf{EHJI}
that the frequencies are unshifted at leading order
since both energy levels in the transition are
increased or decreased by identical amounts.
So,
the standard-model extension
shows no leading-order evidence of Lorentz or CPT violation
in 1S--2S spectroscopy of free \hydrogen\ or \antihydrogen.

One way to proceed
is to consider the next-largest energy-level shifts
in the Lorentz-violating couplings for \hydrogen\ and \antihydrogen.
They are suppressed by a factor of
$\al^2$.
However, the shifts differ in some cases,
and so effects observable in principle
could occur.
One possibility is the
$m_F = 1 \rightarrow m_{F^\prime} = 1$
line, which is shifted by
\beq
\de \nu^H_{1S-2S} \approx - \al^2 b_3^e / 8 \pi
\eeq
relative to the unshifted
$m_F = 0 \rightarrow m_{F^\prime} = 0$
line.
Suppression by a factor
at least of order
$\al^2 \simeq 5\times 10^{-5}$
would also occur in
the proton-antiproton corrections
to the 1S to 2S lines.

Because of these suppression effects in free \hydrogen\ and \antihydrogen\ lines,
it is important to note that
electron-position anomaly-frequency comparisons
in Penning traps \cite{bkr}
are capable of producing tighter bounds
on similar combinations of Lorentz and CPT violating parameters.
Similar g-2 experiments with protons and antiprotons
could in principle also place tighter bounds on proton-antiproton
parameters than would be possible with 1S to 2S spectroscopy.

The advantages of Penning-trap experiments
over spectroscopy in free \hydrogen\ and antihydrogen may
seem counterintuitive given that
g-2 experiments compare gyromagnetic ratios with
fractional resolutions of about $2 \times 10^{-12}$,
as compared to the ultimate fractional resolution
of about $10^{-18}$ in \hydrogen, and perhaps \antihydrogen, experiments.
However, bounds on the coefficients
of the Lorentz- and CPT-violating terms
are determined by absolute,
not relative, resolutions.
In principle,
1S to 2S transitions could reach absolute
resolutions of about one mHz,
as compared with Penning-trap frequencies measured with
about one-Hz resolution.
However, the suppression
in the former by about five orders of magnitude
would effectively give the Penning-trap experiments
a 100-fold
advantage in placing bounds.
These limitations of \hydrogen\ and antihydrogen
in testing Lorentz and CPT symmetry
apply only to the free case
where the transitions maintain the spin direction.
Other considerations apply to
\hydrogen\ and antihydrogen confined by external fields.

\vspace{.4cm}\noindent
{\bf 4. Trapped hydrogen and antihydrogen}\\[.2cm]
Given the technical difficulties of creating antihydrogen atoms,
the advantages of trapping them appear clear.
From the perspective of testing Lorentz and CPT symmetry,
the external trapping fields offer the possibility of separating
different spin states and hence of avoiding the suppression
effects seen in the previous case of free \hydrogen\ and \antihydrogen.
We limit our discussion to the case of
\hydrogen\ or antihydrogen in a constant uniform magnetic field.
This provides an approximation
to the physical environment of
the Ioffe-Pritchard trap \cite{ip},
relevant in the CERN antihydrogen experiments.

In a uniform magnetic field $B$,
the four eigenstates corresponding to any given
principal quantum number $n$
are dependent on the magnetic field.
In increasing energy,
we denote each of the 1S and 2S hyperfine Zeeman levels
by $\ket{a}_n$, $\ket{b}_n$, $\ket{c}_n$, $\ket{d}_n$,
with $n=1$ or $2$,
for both \hydrogen\ and \antihydrogen.
The states for \hydrogen,
expressed in terms of the uncoupled basis
$\ket{m_J,m_I}$ are
\bea
\ket{d}_n &=& \ket{\half, \half} \quad,
\nonumber \\
\ket{c}_n &=& \sin \th_n \ket{-\half,\half} +
\cos \th_n \ket{\half,-\half}
\quad ,
\nonumber\\
\ket{b}_n &=& \ket{-\half, -\half} \quad,
\nonumber \\
\ket{a}_n &=& \cos \th_n \ket{-\half,\half} -
\sin \th_n \ket{\half,-\half}
\quad .
\label{a}
\eea
The mixing angles $\th_n$
are magnetic-field dependent,
and differ for the 1S and 2S states:
\beq
\tan 2 \th_n \approx \fr{(51 {\rm ~mT})}{n^3B}
\quad .
\eeq
The energies of the
$\ket{c}_1$ and $\ket{d}_1$ states
increase with increasing $|B|$,
and so are low-field seekers.
In principle, they remain confined
near the region of minimum field in the trap.
Collisions within the trap,
however,
lead to spin exchange collisions
$
\ket{c}_1 + \ket{c}_1 \rightarrow \ket{b}_1 + \ket{d}_1
$
that tend to decrease the $\ket{c}_1$ population,
and create a predominance of
confined $\ket{d}_1$ states.

The transition between these unmixed-spin states
would appear attractive since it avoids
broadening due to magnetic-field dependence.
A Lorentz or CPT test could be envisaged
involving comparison of the 1S-2S transition
$\ket{d}_1 \rightarrow \ket{d}_2$
in \hydrogen\ and in \antihydrogen.
However,
in both \hydrogen\ and \antihydrogen,
the spin configurations of the
$\ket{d}_1$ and $\ket{d}_2$ states are the same,
so there are again no leading-order shifts
in these frequencies.

A comparison between the
$\ket{c}_1 \rightarrow \ket{c}_2$
transitions in \hydrogen\ and antihydrogen
is of more interest for testing Lorentz or CPT symmetry.
The difference here is that the spin-mixing coefficients
$\sin \th_n$ and $\cos \th_n$
depend on the principal quantum number $n$,
and lead to an unsuppressed shift in the corresponding frequency.
For \hydrogen:
\beq
\de \nu_c^H \approx
-\ka (b_3^e - b_3^p - d_{30}^e m_e
+ d_{30}^p m_p - H_{12}^e + H_{12}^p)/2\pi ~,
\label{nucH}
\eeq
where the spin-mixing function $\ka$ is
\beq
\ka\equiv \cos 2\th_2 - \cos 2\th_1
\quad .
\eeq
It is positive, less than one,
and has maximum value
$\ka \simeq 0.67$
occurring at $B \simeq 0.011$~T.
At this field value the atoms are optimally sensitive
to any Lorentz- or CPT-violating effects.

The frequency $\nu_c^H$ depends on
spatial components of Lorentz-violating couplings
and would therefore vary sidereally due to the rotation
of the Earth.
This effect means it is possible in principle to
detect Lorentz-violating signals with \hydrogen\ alone.
Such clock-comparison tests \cite{qqccexpt} involve comparing
two different transitions,
and have been done with a
\hydrogen\ maser \cite{qqdb}
and other atomic systems \cite{qqlh,qqdp}.
A limitation of these experiments is that they
bound combinations of couplings that violate
Lorentz symmetry,
but do not isolate CPT-violating couplings alone.
The analysis can be complicated because of the
geometry of motion of the laboratory
relative to a suitable inertial reference frame.

In the case of \antihydrogen,
the calculation of
$\de \nu_c^{\overline{H}}$
for the same magnetic field direction and magnitude
involves using the opposite positron and antiproton spins
compared to the electron and proton spins in \hydrogen.
The result is
identical to $\de \nu_c^H$ in Eq.\ \rf{nucH}
except that the signs of $b_3^e$ and $b_3^p$ are reversed:
\beq
\de \nu_c^{\overline H} \approx
-\ka (-b_3^e + b_3^p - d_{30}^e m_e
+ d_{30}^p m_p - H_{12}^e + H_{12}^p)/2\pi ~.
\label{nucAH}
\eeq
The result \rf{nucAH}
can be compared with \rf{nucH},
to give the instantaneous difference
\beq
\De \nu_{1S-2S,c} \equiv \nu_c^H
- \nu_c^{\overline{H}} \approx - \ka (b_3^e - b_3^p)/\pi
\quad .
\label{delcc}
\eeq
This expression involves couplings that are both CPT violating,
and so provides a clean test of CPT symmetry.
It should be noted that the \hydrogen\ and antihydrogen measurements
need to be made at effectively equal times
to ensure that the magnetic field has the same orientation
for both cases.

\vspace{.4cm}\noindent
{\bf 5. Hyperfine Transitions}\\[.2cm]
Hyperfine transitions within the 1S
level of \hydrogen\ can be measured with accuracies
exceeding 1 mHz in masers \cite{ramsey}.
So transitions of this type in trapped \hydrogen\ and antihydrogen
are interesting candidates for
performing tests of Lorentz or CPT symmetry.

To find the energy shifts in the possible transitions
within the 1S level of \hydrogen\ or \antihydrogen,
we may ignore the constant shift of
$a_0^e + a_0^p -c_{00}^e m_e -c_{00}^p m_p$
occurring in each of them,
since this cannot contribute to the difference in energy levels.
The terms of relevance are
\bea
\De E_a^H &\simeq&
\hat\ka (b_3^e - b_3^p - d_{30}^e m_e
+ d_{30}^p m_p - H_{12}^e + H_{12}^p)
\quad ,
\nonumber\\
\De E_b^H &\simeq&
b_3^e + b_3^p - d_{30}^e m_e
- d_{30}^p m_p - H_{12}^e - H_{12}^p
\quad ,
\nonumber\\
\De E_c^H &\simeq& -\De E_a^H
\quad , \\
\De E_d^H &\simeq& - \De E_b^H
\quad ,
\label{abcd}
\eea
where $\hat\ka \equiv \cos2 \th_1$.
Conventionally,
the triplet $\ket b_1$, $\ket c_1$, and $\ket d_1$
is degenerate at zero field.
However,
with the presence of Lorentz- and CPT-violating
couplings,
this degeneracy is broken even at zero field:
equal and opposite energy shifts occur for
$\ket{b}_1$ and $\ket{d}_1$.
In fact,
the difference in the
$\ket{d}_1 \rightarrow \ket{a}_1$
and $\ket{b}_1 \rightarrow \ket{a}_1$
transitions is
unsuppressed and varies sidereally,
\beq
|\De \nu_{d-b}^H| \approx
|b_3^e + b_3^p - d_{30}^e m_e - d_{30}^p m_p
- H_{12}^e - H_{12}^p|/\pi
\quad .
\eeq
In a nonzero magnetic field,
energies are shifted in
all four hyperfine Zeeman levels.
The function $\hat\ka \equiv \cos2 \th_1$
controls the shifts in
$\ket{a}_1$ and $\ket{c}_1$.
As $B$ increases,
so does $\hat\ka$,
reaching $\hat\ka \simeq 1$ when $B \simeq 0.3$ T.

The leading-order effects from
CPT and Lorentz violation
in the standard maser line,
$\ket{c}_1 \rightarrow \ket{a}_1$,
are suppressed.
This is because at the
magnetic field of about $B \lsim 10^{-6}$ Tesla,
the function $\hat\ka$
is approximately $\hat\ka \lsim 10^{-4}$.
Two other candidate transitions for detecting unsuppressed
Lorentz or CPT violation are in the
field-dependent transitions $\ket{d}_1 \rightarrow \ket{a}_1$
and $\ket{b}_1 \rightarrow \ket{a}_1$.
The difference between these two frequencies is
$\De \nu_{d-b}^H$,
and could be used to study
unsuppressed sidereal variations.
One might, however,
expect experimental limitations
such as broadening of the line due to field dependence.

Another possibility would be to measure
radio-frequency transitions between states
within the triplet of hyperfine levels
in \hydrogen\ and \antihydrogen.
This would also offer the possibility of working
at a field-independent point where the limitations
of magnetic-field inhomogeneities can be eliminated
while maintaining sensitivity to Lorentz- and
CPT-violating effects.
An interesting option would be spectroscopy of the
$\ket{d}_1 \rightarrow \ket{c}_1$ transition
at the field-independent point $B \simeq 0.65$~Tesla.
One might expect various experimental challenges
requiring high homogeneity of the field
and extremely low temperatures.
Assuming these issues can be handled,
frequency resolutions of order 1 mHz might be
envisaged.

For this
$\ket{d}_1 \rightarrow \ket{c}_1$ transition
at this field of $B \simeq 0.65$~Tesla,
the electron and proton spins
in state $\ket{c}_1$
interact more strongly with the field
than with each other;
they are highly polarized with
$m_J = 1/2$ and $m_I = - 1/2$.
So, the transition $\ket{d}_1 \rightarrow \ket{c}_1$
is basically a proton spin-flip,
and the frequency shifts are found to be
\beq
\de \nu_{c \rightarrow d}^H  \approx
(-b_3^p + d_{30}^p m_p + H_{12}^p)/\pi
\, , \quad  \quad
\de \nu_{c \rightarrow d}^{\overline{H}} \approx
(b_3^p + d_{30}^p m_p + H_{12}^p)/\pi
\eeq
for \hydrogen\ and
\antihydrogen.
One would thus expect these frequencies
to exhibit sidereal variations.
Another option would be to consider
the instantaneous difference
in these variations,
\beq
\De \nu_{c \rightarrow d} \equiv
\nu_{c \rightarrow d}^H - \nu_{c \rightarrow d}^{\overline{H}}
\approx - 2 b_3^p / \pi
\quad .
\label{nudiff}
\eeq
This comparison could provide a clean and
unsuppressed test of the CPT-violating
coupling $b_3^p$ for the proton.
If, for example,
a frequency resolution of 1 mHz were attained,
this would correspond to an upper bound
of about
\beq
|b_3^p| \lsim 10^{-27} {\textstyle GeV}
\eeq
Compared with estimated bounds in Penning traps,
this is an improvement of about 1000 \cite{bkr}.
It is also more than 10,000 times better
than estimated bounds attainable from 1S-2S transitions.

Clock-comparison experiments are
able to resolve spectral lines to about
1 $\mu$Hz \cite{qqccexpt}.
However,
the coupling expressions they bound \cite{qqkla}
are complex and
isolation of  $b_3^p$ is difficult,
due to the structure of the nuclei involved.
We also note that the experiments discussed here
are sensitive only to spatial components of
Lorentz-violating couplings.
Sensitivity to timelike couplings, like $b_0^e$,
would require a boost,
but would magnify CPT- and Lorentz-violating
effects \cite{qqak}.

\newpage
\vspace{.4cm}\noindent
{\bf 6. Conclusion}\\[.2cm]
The standard-model extension has been used
to analyze various transitions in
the \hydrogen\ and antihydrogen systems,
with the aim of understanding ways in which
Lorentz and CPT symmetry violations may occur.
In free \hydrogen\ and \antihydrogen,
signals for Lorentz and CPT violation
are suppressed by at least one power of
the fine-structure constant.
For trapped \hydrogen\ or \antihydrogen\ atoms,
1S-2S transitions involving the mixed-spin
$\ket{c}$ states, or the spin-flip
$\ket{d}_1 \rightarrow \ket{c}_1$
hyperfine transition could give rise to
unsuppressed signals of
Lorentz and CPT violation.
Such signals would indicate qualitatively new
physics existing in a fundamental theory at the Planck scale.

\vspace{.4cm}\noindent
{\bf Acknowledgments}\\[.2cm]
I thank Robert Bluhm and Alan Kosteleck\'y for their collaboration.
I am grateful for support from the organizers of the Rencontres de Blois XIV
and from Northern Michigan University.

\end{document}